# Bivariate linear mixed models using SAS proc MIXED

Rodolphe Thiébaut[a*], Hélène Jacqmin-Gadda[a], Geneviève Chêne[a], Catherine Leport[b], Daniel Commenges[a]

[a] INSERM Unité 330, ISPED, Université Victor Segalen Bordeaux II, 146, rue Léo Saignat 33076, Bordeaux Cedex, France

[b] Hôpital Bichat Claude Bernard, Paris, France

## Abstract

Bivariate linear mixed models are useful when analyzing longitudinal data of two associated markers. In this paper, we present a bivariate linear mixed model including random effects or first-order auto-regressive process and independent measurement error for both markers. Codes and tricks to fit these models using SAS Proc MIXED are provided. Limitations of this program are discussed and an example in the field of HIV infection is shown. Despite some limitations, SAS Proc MIXED is a useful tool that may be easily extendable to multivariate response in longitudinal studies.

* Corresponding author. Tel: +33 5 57 57 45 21 Fax: +33 5 56 24 00 81

Email: rodolphe.thiebaut@isped.u-bordeaux2.fr

# 1. Introduction

Longitudinal data are often collected in epidemiological studies, especially to study the evolution of biomedical markers. Thus, linear mixed models [1], recently available in standard statistical packages [2, 3], are increasingly used to take into account all available information and deal with the intra-subject correlation.

When several markers are measured repeatedly, longitudinal multivariate models could be used, like in econometrics. However, this extension of univariate models is rarely used in biomedicine although it could be useful to study the joint evolution of biomarkers. Into example, in HIV infection, several markers are available to measure the quantity of virus (plasma viral load noted HIV RNA), the status of immune system (CD4+ T lymphocytes which are a specific target of the virus, CD8+ T lymphocytes) or the inflammation process ($\beta_2$ microglobuline). These markers are associated as the infection measured by HIV RNA induces inflammation and the destruction of immune cells. Several authors have developed methods to fit evolution of CD4 and CD8 cells [4] or CD4 and $\beta_2$ microglobuline [5]. Amrick Shah et al. [4] used an EM algorithm to fit a bivariate linear random effects model. Sy et al [5] used the Fisher scoring method to fit a bivariate linear random effects model including an integrated Orstein-Uhlenbeck process (IOU). IOU is a stochastic process that includes Brownian motion as special limiting case.

Their programs were implemented using IML module of SAS Software [6]. However, their flexibility is not sufficient to allow a large use by researchers not familiar with IML. Also, the EM algorithm chosen is slow.

In this paper, we propose some tricks to use SAS MIXED procedure in order to fit multivariate linear mixed models to multivariate longitudinal gaussian data. SAS MIXED procedure uses Newton-Raphson algorithm known to be faster than the EM algorithm [7]. In section 2 and 3, we present bivariate linear mixed models and the code used in SAS to fit these models. In section 4, we apply these models to study the joint evolution of HIV RNA and CD4+ T lymphocytes in a cohort of HIV-1 infected patients (APROCO) treated with highly active antiretroviral treatment.

## 2. Model for bivariate longitudinal gaussian data

We define a general bivariate linear mixed model including a random component, a first order auto-regressive process and an independent error.

Let $Y_i = \begin{bmatrix} Y_i^1 \\ Y_i^2 \end{bmatrix}$, the response vector for the subject i, with $Y_i^k$ the $n_i^k$-vector of measurements of the marker k (k=1, 2) with $n_i^1 = n_i^2 = n_i$. If the two markers are independent, we can use the two following models:

$$\begin{cases} Y_i^1 = X_i^1 \beta^1 + Z_i^1 \gamma_i^1 + W_i^1 + \varepsilon_i^1 \\ Y_i^2 = X_i^2 \beta^2 + Z_i^2 \gamma_i^2 + W_i^2 + \varepsilon_i^2 \end{cases} \text{where} \begin{array}{ll} \varepsilon_i^1 \sim N(0, \sigma_{\varepsilon^1}^2 I_{n_i}) & \varepsilon_i^2 \sim N(0, \sigma_{\varepsilon^2}^2 I_{n_i}) \\ \gamma_i^1 \sim N(0, G^1) & \text{and} \quad \gamma_i^2 \sim N(0, G^2) \\ W_i^1 \sim N(0, R_i^1) & W_i^2 \sim N(0, R_i^2) \end{array}$$

where $X_i^k$ is a $n_i \times p^k$ design matrix, $\beta^k$ is a $p^k$-vector of fixed effects, $Z_i^k$ is a $n_i \times q^k$ design matrix which is usually a subset of $X_i^k$, $\gamma_i^k$ is a $q^k$-vector of individual random effects with $q^k \leq p^k$. $W_i^k$ is a vector of realization of a first order auto-regressive process $w_i^k(t)$ with covariance given by $R_i^k(s,t) = \sigma_{w^k}^2 e^{\lambda^k |t-s|}$ and $I_{n_i}$ is a $n_i \times n_i$ identity matrix.

To take into account correlation between both markers, one could use the following bivariate linear mixed model:

$$Y_i = X_i \beta + Z_i \gamma_i + W_i + \varepsilon_i \text{ with } \begin{cases} \varepsilon_i \sim N(0, \Sigma_i) \\ W_i \sim N(0, R_i) \\ \gamma_i \sim N(0, G) \end{cases}$$

where $X_i = \begin{bmatrix} X_i^1 & 0 \\ 0 & X_i^2 \end{bmatrix}$, $\beta = \begin{bmatrix} \beta_1 \\ \beta_2 \end{bmatrix}$, $Z_i = \begin{bmatrix} Z_i^1 & 0 \\ 0 & Z_i^2 \end{bmatrix}$, $\gamma_i = \begin{bmatrix} \gamma_i^1 \\ \gamma_i^2 \end{bmatrix}$ and $W_i = \begin{bmatrix} W_i^1 \\ W_i^2 \end{bmatrix}$ is a $2n_i$-

vector of realization of a bivariate first order auto-regressive process $w_i(t) = \begin{bmatrix} w_i^1(t) \\ w_i^2(t) \end{bmatrix}$ and

$\varepsilon_i = \begin{bmatrix} \varepsilon_i^1 \\ \varepsilon_i^2 \end{bmatrix}$ represents independent measurement errors.

The covariance matrix of measurement errors is defined by $\Sigma_i = \Sigma \otimes I_{n_i}$ and $\Sigma = \begin{bmatrix} \sigma_{\varepsilon^1}^2 & 0 \\ 0 & \sigma_{\varepsilon^2}^2 \end{bmatrix}$

(the symbol $\otimes$ represents the Kronecker product). The covariance function of the bivariate auto-regressive process $w_i(t) = \begin{bmatrix} w_i^1(t) \\ w_i^2(t) \end{bmatrix}$ is given by $R_i(s,t) = C \times e^{B|t-s|}$ with

$C = \begin{bmatrix} \sigma_{w^1}^2 & \sigma_{w^1 w^2} \\ \sigma_{w^1 w^2} & \sigma_{w^2}^2 \end{bmatrix}$ is the process covariance matrix at $t = s$ and B is a $2 \times 2$ matrix such that

(i) the eigenvalues of B have negative real parts, and (ii) C and $D = -(CB + B'C)$ are positive

definite symmetric [5]. The covariance matrix of random effects is the matrix

$G = \begin{bmatrix} G^1 & G^{12} \\ G^{12} & G^2 \end{bmatrix}$. With the assumption that $\gamma_i$, $W_i$ and $\varepsilon_i$ are mutually independent, it is obvious that $\text{var}(Y_i) = V_i = Z_i G_i Z_i^T + R_i + \Sigma_i$.

## 3. Models using Proc MIXED of SAS software

### *3.1 Random effects*
As described in the documentation [3], multivariate random effects models can be fitted using the statement random and an indicator variable for each marker to define $Y_i^k$, $X_i^k$ and $Z_i^k$.

To add an independent error for each response variable in a multivariate random effect model,

one must use the repeated statement with the option *GROUP(VAR)* where VAR is a binary variable indicating the response variable concerned (VAR=0 for $Y^1$ and VAR=1 for $Y^2$ into example). This option allows estimation of heterogeneous covariance structure, i.e. the variances of the measurement errors are different for each response variable.

An example of SAS code for a bivariate random effect model with random intercept and random slopes is:

```
Proc mixed data=BIV;
class CEN_PAT VAR;
model Y=VAR VAR*T;
random VAR VAR*T /type=UN sub=CEN_PAT;
repeated /type=VC grp=VAR sub=CEN_PAT;
run ;
```

where CEN_PAT is a single identification number of each patient and T is time. In this example, we have $p^1 = p^2 = q^1 = q^2 = 2$ and $X_{ij}^1 = X_{ij}^2 = Z_{ij}^1 = Z_{ij}^2 = \begin{bmatrix} 1 & t_{ij} \end{bmatrix}$ for the measurement j of the subject i. Note that the two markers are independent if $G = \begin{bmatrix} G^1 & 0 \\ 0 & G^2 \end{bmatrix}$.

## 3.2 First order auto-regressive process

In the repeated statement SAS provides the possibility to fit bivariate models using a Kronecker product notation [8]. For instance, in the bivariate case with 3 repeated measures, the option *type=UN@AR(1)* in the statement *repeated* assumes that the covariance matrix has the following structure: $\begin{bmatrix} \sigma_{w^1}^2 & \sigma_{w^1w^2} \\ \sigma_{w^1w^2} & \sigma_{w^2}^2 \end{bmatrix} \otimes \begin{bmatrix} 1 & \rho & \rho^2 \\ \rho & 1 & \rho \\ \rho^2 & \rho & 1 \end{bmatrix}$. Compared with the general bivariate auto-regressive process defined in the previous section, this structure has two important limitations. First, the covariance structure is a first order auto-regressive process for discrete data and assumes the measures are equally spaced for all subjects and for the two markers. In the univariate case, a continuous time AR(1) model, which allows non equally spaced measures, may be fitted using the structure *SP(POW)* but this structure is not available for multivariate models. The second limitation is that the SAS program allows to estimate only one correlation parameter $(\rho)$ for the 'bivariate process' rather than a matrix B. Thus, using this formulation, one assumes that the intra-marker correlation is the same for the two markers, i.e. $Corr\left(w_i^1(s), w_i^1(t)\right) = Corr\left(w_i^2(s), w_i^2(t)\right) = \rho^{|t-s|}$. Moreover, one assumes that inter-marker correlation is proportional to the intra-marker correlation, i.e. $Corr\left(w_i^1(s), w_i^2(t)\right) = \frac{\sigma_{w^1 w^2}}{\sigma_{w^1} \sigma_{w^2}} \cdot \rho^{|t-s|}$. Both markers are independent if the covariance matrix has

the form $\begin{bmatrix} \sigma_{w^1}^2 & 0 \\ 0 & \sigma_{w^2}^2 \end{bmatrix} \otimes \begin{bmatrix} 1 & \rho & \rho^2 \\ \rho & 1 & \rho \\ \rho^2 & \rho & 1 \end{bmatrix}$.

To add an independent measurement error for both markers, one must use the option *LOCAL(EXP <effects>)* which produces exponential local effects, *<effects>=VAR* being still the indicator variable of response variable. These local effects have the form $\sigma_\varepsilon^2 diag[\exp(U\delta)]$ where **U** is a full-rank design matrix. PROC MIXED constructs **U** in terms of 1s and −1s for a classification effect and estimates $\delta$.

An example of SAS code to fit a bivariate first-order auto-regressive model is:

```
Proc mixed data=BIV;
class CEN_PAT VAR;
model Y=VAR VAR*T;
repeated VAR /type=UN@AR(1) local=exp(VAR) sub=CEN_PAT;
run ;
```

where T is the time as a continuous variable and VAR is the indicator variable.

The SAS output contains the following covariance parameters estimates: 'VAR UN(x,y)' which correspond to the matrix containing covariance parameter of the auto-regressive process $\begin{bmatrix} \sigma_{w^1}^2 & \sigma_{w^1 w^2} \\ \sigma_{w^1 w^2} & \sigma_{w^2}^2 \end{bmatrix}$, 'EXP VAR' which is the local effect parameter $(\delta)$, 'Residual' that we noted $r$ and a parameter called 'AR(1)'. From this output, the parameters of the model could be calculated as: $\sigma_{\varepsilon^1}^2 = r \times e^{\delta}$, $\sigma_{\varepsilon^2}^2 = r \times e^{-\delta}$ and $\rho = \dfrac{AR(1)}{r}$.

### 3.3 Incomplete data

Likelihood based inference used by Proc MIXED is valid whenever the mechanism of missing data is ignorable, that is MAR (Missing at Random), i.e. the availability of the measurement do not depend on the true value of the marker at the same time, and the parameters describing the non-response mechanism are distinct from the model parameters [9]. However, using an auto-regressive process, one must be careful when missing data occur. By default, a dropout mechanism is assumed to be responsible of missing data by MIXED procedure: all missing data are considered to occur after the last observed measurement. To take into account for intermittent missing data (one observation missing between two observed), a class variable must be used in the *repeated* statement indicating the order of observations within a subject. In the following example, the class variable is a copy of the variable time, named 'Tbis' :

```
Proc mixed data=BIV;
class CEN_PAT VAR Tbis;
model Y=VAR VAR*T;
repeated VAR Tbis /type=UN@AR(1) local=exp(VAR) sub=CEN_PAT;
run ;
```

When the measurements of the two markers never occur at the same time because of a design consideration, auto-regressive process can not be used unlike random effects model.

## 4. Application

### 4.1 The APROCO Cohort

The APROCO (ANRS-EP11) cohort is a prospective observational cohort ongoing in 47 clinical centres in France. A total of 1,281 HIV-1-infected patients were enrolled from May 1997 to June 1999 at the initiation of their first highly active antiretroviral therapy containing a protease inhibitor. Standardised clinical and biological data including CD4+ cell counts measurements and plasma HIV RNA quantification were collected at baseline (M0), one month later (M1) and every 4 months (M4-M24) thereafter. In order to ensure sufficient available information, only a sub-sample of patients having both plasma HIV RNA and CD4+ cell counts measurements at M0 and at least two measurements thereafter were included in the analyses. The first measurement after baseline (at one month) was deleted to provide a data set with equally spaced measures. Follow-up data were included until the 24$^{th}$ month; thus patients had a maximum of 7 measures. The study population and evolution of virological response were described elsewhere [10]. Available information at each study time and description of the evolution of both markers were presented in table 1 and figure 1.

## 4.2 Modeling

To assure normality and homoskedasticity of residuals distribution, variable response was the change in value of marker at time t since the initial visit, i.e. $Y_i^1(t) = \log_{10} HIVRNA(t) - \log_{10} HIVRNA(0)$ and $Y_i^2(t) = CD_4(t) - CD_4(0)$.

Fixed effects included a change of slope intensity at time 4 months as suggested in figure 1. Note that we did not include intercept because $Y_i^1(0) = Y_i^2(0) = 0 \quad \forall i$.

We compared 4 models providing two forms of covariance structure (random effects or auto-regressive process) in two formulations (univariate or bivariate). Univariate and bivariate random effect models were compared using likelihood ratio test as both models were nested. The bivariate model had only four covariance parameters in addition. Comparison of random effects versus auto-regressive process were performed using AIC criteria [11]. A general model including random slopes and a bivariate first order auto-regressive process did not converge as reported in univariate cases by others (see [12] for example).

The model including two random slopes and a measurement error for each marker was:

$$\begin{cases} Y_i^1 = \beta_1^1(t_i \wedge \tau) + \beta_2^1(t_i - \tau)I_{t_i \geq \tau} + \gamma_{1i}^1(t_i \wedge \tau) + \gamma_{2i}^1(t_i - \tau)I_{t_i \geq \tau} + \varepsilon_i^1 \\ Y_i^2 = \beta_1^2(t_i \wedge \tau) + \beta_2^2(t_i - \tau)I_{t_i \geq \tau} + \gamma_{1i}^2(t_i \wedge \tau) + \gamma_{2i}^2(t_i - \tau)I_{t_i \geq \tau} + \varepsilon_i^2 \end{cases}$$

where $\beta_1^k$ is the first slope before the time $\tau = 4 \; months$, $\beta_2^k$ is the second slope after the time $\tau$ and $t_i \wedge \tau$ represents the minimum between $t_i$ and $\tau$.

$$\text{Moreover,} \begin{pmatrix} \gamma_{1i}^1 \\ \gamma_{2i}^1 \\ \gamma_{1i}^2 \\ \gamma_{2i}^2 \end{pmatrix} \sim N \left( \begin{pmatrix} 0 \\ 0 \\ 0 \\ 0 \end{pmatrix}, \begin{pmatrix} \sigma_{\gamma_{1i}^1}^2 & \sigma_{\gamma_{1i}^1 \gamma_{2i}^1} & \sigma_{\gamma_{1i}^1 \gamma_{1i}^2} & \sigma_{\gamma_{1i}^1 \gamma_{2i}^2} \\ \sigma_{\gamma_{1i}^1 \gamma_{2i}^1} & \sigma_{\gamma_{2i}^1}^2 & \sigma_{\gamma_{2i}^1 \gamma_{1i}^2} & \sigma_{\gamma_{2i}^1 \gamma_{2i}^2} \\ \sigma_{\gamma_{1i}^1 \gamma_{1i}^2} & \sigma_{\gamma_{2i}^1 \gamma_{1i}^2} & \sigma_{\gamma_{1i}^2}^2 & \sigma_{\gamma_{1i}^2 \gamma_{2i}^2} \\ \sigma_{\gamma_{1i}^1 \gamma_{2i}^2} & \sigma_{\gamma_{2i}^1 \gamma_{2i}^2} & \sigma_{\gamma_{1i}^2 \gamma_{2i}^2} & \sigma_{\gamma_{2i}^2}^2 \end{pmatrix} \right)$$

The model including an auto-regressive process and a measurement error was:

$$\begin{cases} Y_i^1 = \beta_1^1 t_i \wedge \tau + \beta_2^1(t_i - \tau)I_{t_i \geq \tau} + W_i^1 + \varepsilon_i^1 \\ Y_i^2 = \beta_1^2 t_i \wedge \tau + \beta_2^2(t_i - \tau)I_{t_i \geq \tau} + W_i^2 + \varepsilon_i^2 \end{cases} \text{ where } \begin{bmatrix} W_i^1 \\ W_i^2 \end{bmatrix} \sim N\left( \begin{pmatrix} 0 \\ 0 \end{pmatrix}, R_i \right)$$

where $R_i = \begin{bmatrix} \sigma_{w^1}^2 & \sigma_{w^1 w^2} \\ \sigma_{w^1 w^2} & \sigma_{w^2}^2 \end{bmatrix} \otimes \begin{bmatrix} 1 & \rho & \ldots & \rho^7 \\ \rho & 1 & \rho & \ldots \\ \ldots & \rho & \ldots & \rho \\ \rho^7 & \ldots & \rho & 1 \end{bmatrix}$.

### *4.3 SAS programming*

The initial data set had the following presentation :

| CEN_PAT | CD4 | RNA | T |
|---|---|---|---|
| 1001 | 166 | -3.02635 | 4 |
| 1001 | 147 | -1.96563 | 8 |
| 1001 | 171 | -1.42426 | 12 |
| 1001 | 355 | -1.07208 | 16 |
| 1001 | 223 | -3.38035 | 20 |
| 1001 | 52 | -2.08382 | 24 |
| 1002 | -14 | -2.84515 | 4 |
| 1002 | -123 | -2.84515 | 8 |

With CEN_PAT being the patient number, CD4 the difference in CD4 cell count since baseline, RNA the difference in HIV RNA since baseline and T the date of measurement in months. The change in slope intensity at 4 months was computed using a data step:

```
Data file; set file;
if T<4 then do ; T1=T;T2=0; end ;
if T ge 4 then do; T1=CP;T2=T-4;
end ;
```

Then, the structure of input data was transformed to allow bivariate modeling. Mainly, it consists in the integration of CD4 and HIV RNA in the same vector (Y here) and an indicator variable (VAR here)

```
Data var0; set file;
VAR=0;Y=RNA;
keep CEN_PAT Y VAR T T1 T2;

Data var1;set file;
VAR=1;Y=CD4;
keep CEN_PAT Y VAR T T1 T2;

Data biv ;set var0 var1 ;
run ;
```

Thus, a bivariate random effect model was fitted using the code described below.

```
Proc mixed data=BIV CL;
class CEN_PAT VAR;
model Y=VAR*T1 VAR*T2/noint s;
random VAR*T1 VAR*T2/type=UN sub=CEN_PAT G GCORR;
repeated /type=VC grp=VAR sub=CEN_PAT;
run ;
```

The option "CL" requests confidence limits for the covariance parameter estimates. A Satterthwaite approximation is used to construct limits for all parameters that have a default lower boundary constraint of zero. In the statement model, the option "noint" was used to avoid the inclusion of intercepts and "s" to obtain solution for fixed effects.

In the same way, a bivariate model with an auto-regressive process and separate measurement errors was fitted using the following code:

```
Proc mixed data=BIV CL;
class CEN_PAT VAR T;
model Y=VAR*T1 VAR*T2 /noint s;
repeated VAR T/type=UN@AR(1) local=exp(VAR) sub=CEN_PAT;
run ;
```

## *4.4 Results*

The bivariate random effects model was significantly better than two separate univariate random effects models (-25194 vs. -25307, likelihood ratio = 226 with 4 degrees of freedom, $p<10^{-4}$, table 2) showing a strong association between the two markers. The bivariate random effect model allows to estimate the correlation matrix between individual slopes for each marker. In this correlation matrix, every element was significantly ($p < 0.05$) different from 1 (table 3). Briefly, the highest correlations were between the slopes of the two markers at the same period: ($\rho(\beta_1^{CD4}, \beta_1^{HIVRNA}) = -0.41$ before 4 months and $\rho(\beta_2^{CD4}, \beta_2^{HIVRNA}) = -0.60$ after 4 months). These results were expected because of biological relation between the two markers. Moreover, the second slope of CD4 cell count was highly correlated to the first slope of the same marker $\rho(\beta_1^{CD4}, \beta_2^{CD4}) = 0.37$.

The bivariate model including a bivariate auto-regressive process was better than the bivariate random effects model despite the restrictive assumption that the two intra-marker correlations are equal (AIC 50386 vs. 50646).

Output obtained with the model including a first order auto-regressive process provide estimations of $\sigma_{w^1}^2 = 1.54$, $\sigma_{w^2}^2 = 195$ and $\sigma_{w^1 w^2} = -7.00$, significantly different from 0 (Wald test, p<10$^{-4}$). This last result underlines the relationship between the two markers. The parameter $\rho = \frac{3.11}{3.42} = 0.91$ is the correlation between two consecutive measures of CD4 cell count or HIV RNA. Variances of measurement error are calculated as: $\sigma_{\varepsilon^1}^2 = 3.42 \, e^{3.11} = 77.00$ and $\sigma_{\varepsilon^2}^2 = 3.42 \, e^{-3.11} = 0.15$.

Thus, the relationship between the two makers was underlined by the correlation between the markers at each period and the improvement of likelihood of the bivariate model compared to two univariate models. Bivariate random effect model offers a direct interpretation of the relationship between the markers without assumption on the dependence of one marker in relation to the other.

## 5. Conclusion

Bivariate models are useful for longitudinal data in biomedical research and can be computed using standard statistical package like the SAS system. Moreover, the efficiency of the procedure MIXED, which allows quick convergence, should be underlined. However, there are some limitations inherent in the identical intra-marker correlations or in the assumption of constant period between two measurements for the first order auto-regressive covariance structure implemented in the SAS system. Finally, although the number of parameters would dramatically increase, particularly in the case of multivariate random effect model, bivariate models are easily extendable to multivariate models with more than two dependent variables.

Table 1. Measures of CD4 cell count and HIV RNA during follow-up. APROCO Study (N=988).

| Month | Change in CD4 cell count / mm$^3$ from baseline | | | Change in log$_{10}$ copies/ml HIV RNA from baseline | | |
|---|---|---|---|---|---|---|
| | N | Mean | SD | N | Mean | SD |
| 4 | 988 | 97 | 130 | 988 | -1.95 | 1.20 |
| 8 | 935 | 126 | 147 | 919 | -2.01 | 1.27 |
| 12 | 901 | 153 | 169 | 894 | -2.04 | 1.34 |
| 16 | 823 | 176 | 180 | 813 | -2.03 | 1.35 |
| 20 | 708 | 192 | 190 | 703 | -2.00 | 1.37 |
| 24 | 534 | 201 | 196 | 530 | -1.93 | 1.37 |

Table 2. Likelihood of models according to the type of covariance matrix. APROCO Study (N=988).

|  | Log Likelihood | No. of parameters | AIC |
|---|---|---|---|
| Univariate model with two random slopes | -25307 | 12 | 50638 |
| Bivariate model with two random slopes | -25194 | 16 | 50420 |
| Univariate model with AR(1) | -25313 | 10 | 50646 |
| Bivariate model with AR(1) | -25183 | 10 | 50386 |

$AIC = (-2 \log likelihood) + 2 (No. of\ parameters)$  AR(1) : First order auto-regressive process

Table 3. Estimated correlation matrix of the bivariate model including two random slopes. APROCO Study (N=988).

|  | First slope of HIV RNA | Second slope of HIV RNA | First slope of CD4+ | Second slope of CD4+ |
|---|---|---|---|---|
| First slope of HIV RNA | 1 | | | |
| Second slope of HIV RNA | -0.10 | 1 | | |
| First slope of CD4+ | -0.41 | 0.13 | 1 | |
| Second slope of CD4+ | -0.16 | -0.60 | 0.37 | 1 |

Figure 1. Mean change in observed HIV RNA and CD4+ cell count (95% confidence interval) after initiation of an antiretroviral treatment containing a protease inhibitor. APROCO study (N=988).

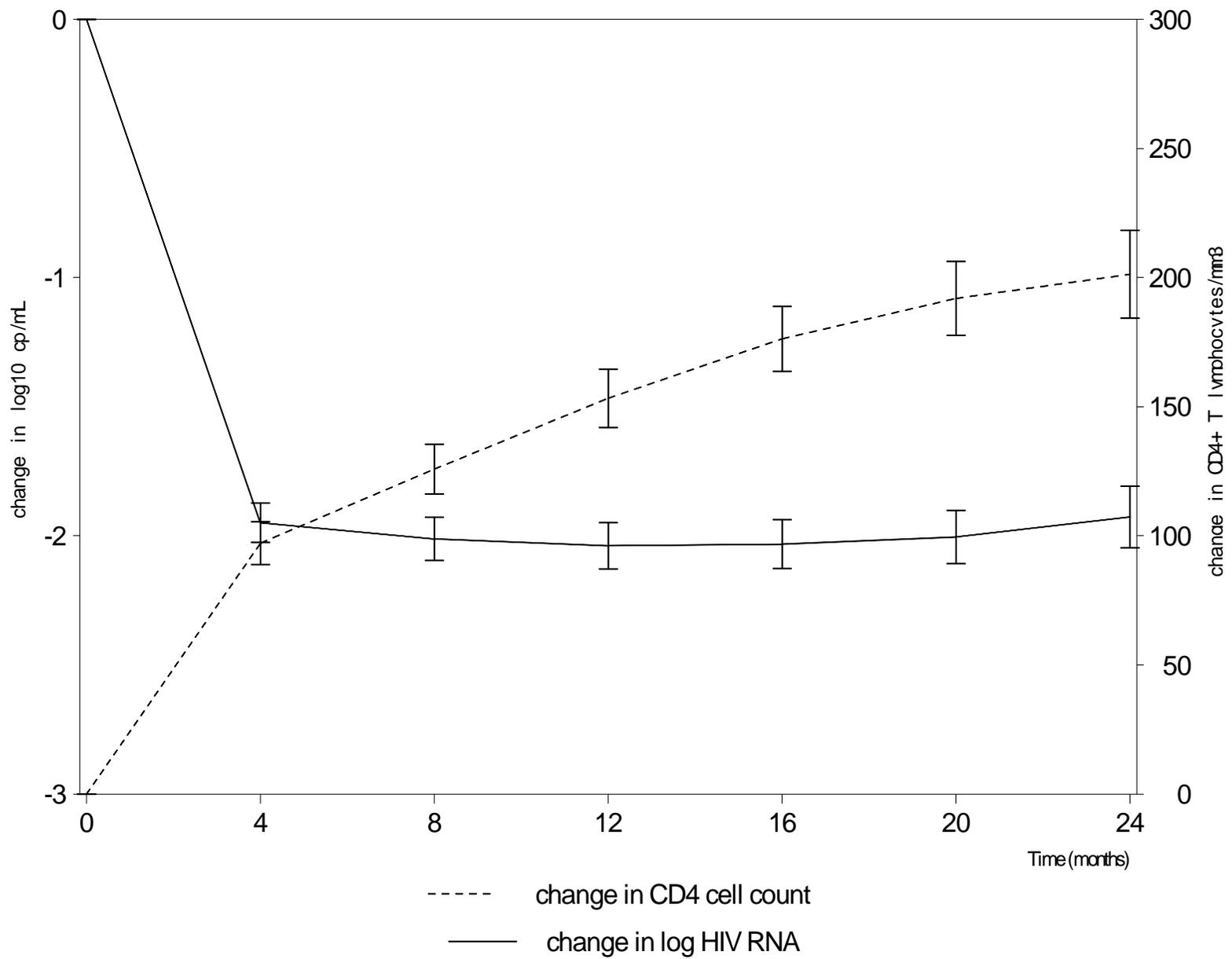